%% file: main.tex
\documentclass[journal = jctcce, manuscript=article]{achemso}
\usepackage{ulem}

\newcommand{\br}{\boldsymbol{\textbf{r}}}

\newcommand{\bR}{\boldsymbol{\textbf{R}}}

\newcommand{\dr}{\,d\br}

\newcommand{\rhoWF}{\rho^{\scriptsize{\text{WF}}}}
\newcommand{\rhoKS}{\rho^{\scriptsize{\text{KS}}}}

\newcommand{\vxc}{v_{\rm{xc}}}

\usepackage{doi}
\usepackage{mathtools}
\usepackage{hyperref}
\hypersetup{
    colorlinks=true,
    linkcolor=magenta,
    filecolor=magenta,      
    urlcolor=cyan,
    }

\usepackage{subfiles}
\usepackage{graphicx}
\usepackage{graphics}
\usepackage{multirow}
\usepackage{amsxtra}
\usepackage{stmaryrd}
\usepackage{mathrsfs}
\usepackage{caption}
\usepackage{subcaption}
\usepackage{epsfig}
\usepackage{rotating}
\usepackage{setspace}
\usepackage{float}
\usepackage{amsfonts}
\usepackage{amsbsy}
\usepackage{amscd}
\usepackage{amsthm}
\usepackage{relsize}
\usepackage{color}
\usepackage{tabularx}
\usepackage{caption}
\usepackage{sidecap}
\usepackage{dcolumn}
\usepackage{footnote}
\usepackage{algorithm}
\usepackage{algorithmic}
\usepackage{breqn}

\definecolor{hellgruen}{rgb}{0.2,0.7,0.2}

\newcommand{\cn}{\color{black}}

\newcolumntype{M}[1]{>{\centering\arraybackslash}m{#1}}
\newcolumntype{N}{@{}m{0pt}@{}}
\title{Accelerating inverse Kohn-Sham calculations using reduced density matrices}

\author{Bikash Kanungo}
\affiliation{Department of Mechanical Engineering, University of Michigan, Ann Arbor, Michigan 48109, USA}
\author{Soumi Tribedi}
\affiliation{Department of Chemistry, University of Michigan, Ann Arbor, Michigan 48109, USA}
\author{Paul M. Zimmerman}
\affiliation{Department of Chemistry, University of Michigan, Ann Arbor, Michigan 48109, USA}
\author{Vikram Gavini}
\affiliation{Department of Mechanical Engineering, University of Michigan, Ann Arbor, Michigan 48109, USA}
\alsoaffiliation{Department of Materials Science and Engineering, University of Michigan, Ann Arbor, Michigan 48109, USA}
\email{vikramg@umich.edu}

\begin{document}
\begin{abstract}
The Ryabinkin–Kohut–Staroverov (RKS) and Kanungo-Zimmerman-Gavini (KZG) methods offer two approaches to find exchange-correlation (XC) potentials from ground state densities. The RKS method utilizes the one- and two-particle reduced density matrices to alleviate any numerical artifacts stemming from a finite basis (e.g., Gaussian- or Slater-type orbitals). The KZG approach relies solely on the density to find the XC potential, by combining a systematically convergent finite-element basis with appropriate asymptotic correction on the target density. The RKS method, being designed for a finite basis, offers computational efficiency. The KZG method, using a complete basis, provides higher accuracy. In this work, we combine both the methods to simultaneously afford accuracy and efficiency. In particular, we use the RKS solution as initial guess to the KZG method to attain a significant $3-11\times$ speedup. This work also presents a direct comparison of the XC potentials from the RKS and the KZG method and their relative accuracy on various weakly and strongly correlated molecules, using their ground state solutions from accurate configuration interaction calculations solved in a Slater orbital basis.    
\end{abstract}

\input{intro}

\input{methods}

\input{results}

\input{conclusion}

\section*{Acknowledgements} 
We gratefully acknowledge DOE grant DE-SC0022241 which supported this study. This research used resources of the NERSC Center, a DOE Office of Science User Facility supported by the Office of Science of the U.S. Department of Energy under Contract No. DE-AC02-05CH11231. We acknowledge the support of DURIP grant W911NF1810242, which also provided computational resources for this work.

\input{bibitem_main}

\newpage
\begin{figure}
    \centering
    \includegraphics[scale=1.0]{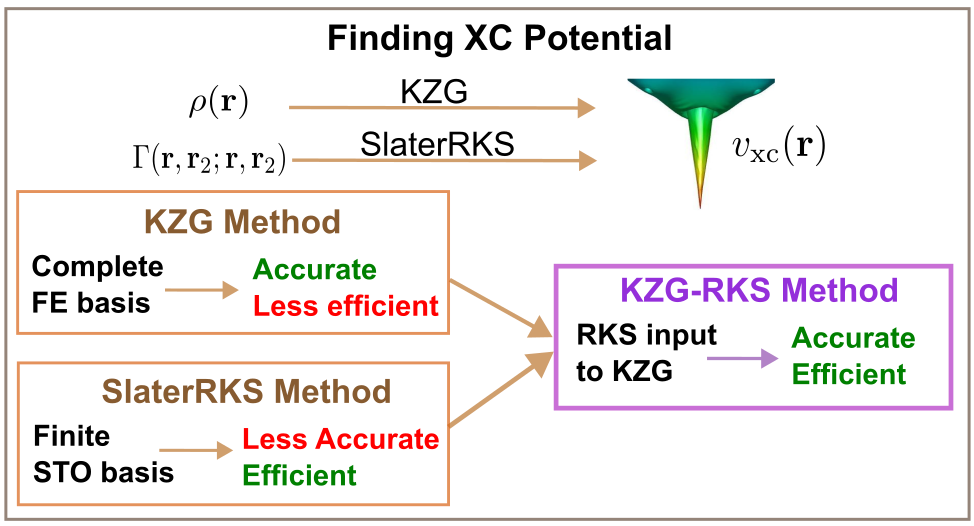}
    \caption*{TOC graphic}
    \label{fig:ToC}
\end{figure}

\pagebreak
\begin{center}
\textbf{\Large Supplemental Material}
\end{center}

\setcounter{equation}{0}
\setcounter{figure}{0}
\setcounter{table}{0}
\setcounter{page}{1}
\setcounter{section}{0}
\makeatletter
\renewcommand{\theequation}{S\arabic{equation}}
\renewcommand{\thefigure}{S\arabic{figure}}
\renewcommand{\thetable}{S\arabic{table}}
\renewcommand{\bibnumfmt}[1]{[S#1]}
\renewcommand{\citenumfont}[1]{S#1}
\renewcommand{\thesection}{S\arabic{section}}
\renewcommand{\thepage}{S-\Roman{page}}

\begin{table}[h!]
    \caption{Coordinates for all benchmark systems (in Angstrom).}
    \centering
    \begin{tabular}{M{0.2\columnwidth}M{0.1\columnwidth}M{0.1\columnwidth}M{0.1\columnwidth}M{0.1\columnwidth}}\hline
     $\text{H}_2(eq)$ & & & & \\ \hline
     & H & $0$ & $0$ & $0$\\  
     & H & $0$ & $0$ & $0.74$\\ \hline 
     $\text{H}_2(2eq)$ & & & & \\ \hline
     & H & $0$ & $0$ & $0$\\  
     & H & $0$ & $0$ & $1.5$\\ \hline  
     $\text{H}_2(d)$ & & & & \\ \hline
     & H & $0$ & $0$ & $0$\\  
     & H & $0$ & $0$ & $4.0$\\ \hline 
     LiH & & & & \\ \hline
     & Li	& $0$ & $0$ & $0$ \\
     & H	& $0$ & $0$	& $1.5949$ \\ \hline
     $\text{H}_2\text{O}$ & & & & \\ \hline
     & O & $0$ & $0$ & $0$\\
     & H & $0$ & $0$ & $1.0$\\  
     & H & $0.96982$	& $0$ & $-0.24380$ \\ \hline
     CH$_2$ - singlet & & & & \\ \hline			
     & C	& $0$ & $0$	& $0$ \\ 
     & H	& $0$	& $0$	& $1.10664$ \\
     & H	& $0$	& $1.08092$	& $-0.23719$ \\ \hline
    \end{tabular}
    \begin{tabular}{M{0.6\columnwidth}}
     Continued on next page \\ \hline
    \end{tabular}
\end{table}
\addtocounter{table}{-1}
\newpage

\begin{table}[h!]
    \caption{\emph{Continuing from previous page}}
    \centering
    \begin{tabular}{M{0.1\columnwidth}M{0.1\columnwidth}M{0.1\columnwidth}M{0.1\columnwidth}M{0.1\columnwidth}}\hline
    $\text{C}_2\text{H}_{4}$ & & & & \\ \hline   
    & C	& $0$ & $0$	& $0.6695$ \\ 
    & C	& $0$ & $0$	& $-0.6695$ \\
    & H	& $0$ & $0.9289$ & $1.2321$ \\
    & H	& $0$ & $-0.9289$ & $1.2321$ \\
    & H	& $0$ & $0.9289$ & $-1.2321$ \\ 
    & H	& $0$ & $-0.9289$ & $-1.2321$ \\ \hline
    
    $\text{C}_2\text{H}_{4}$(2eq) & & & & \\ \hline   
    & C	& $0$ & $0$	& $1.339$ \\ 
    & C	& $0$ & $0$	& $-1.339$ \\
    & H	& $0$ & $0.9289$ & $1.9016$ \\
    & H	& $0$ & $-0.9289$ & $1.9016$ \\
    & H	& $0$ & $0.9289$ & $-1.9016$ \\ 
    & H	& $0$ & $-0.9289$ & $-1.9016$ \\ \hline
    \end{tabular}
\end{table}

\end{document}

%% file: intro.tex
\section{Introduction}
Density functional theory (DFT)~\cite{Becke2014} has been the workhorse for electronic structure calculations for decades. It  relies on the Hohenberg-Kohn theorem~\cite{Hohenberg1964} and the Kohn-Sham \textit{ansatz}~\cite{Kohn1965} to formally reduce the many-electron Schr\"odinger equation to an equivalent problem of non-interacting electrons in an effective mean-field governed by the ground-state electron density. Although exact in principle, in practice DFT requires approximation to the unknown exchange-correlation (XC) functional that encapsulates the quantum many-electron interactions into a mean-field of the density. Despite decades of development, fundamental deficiencies, such as delocalization~\cite{Bryenton2023}, static correlation error~\cite{Cohen2008}, inability to capture strong correlations~\cite{Cohen2012}, etc., persists in all existing XC approximations. 
The \textit{inverse} DFT problem, that maps the density (say from accurate wavefunction-based calculation) to its XC potential, can be instrumental in developing accurate functionals via machine-learning~\cite{Schmidt2019, Zhou2019, Nagai2020} as well as to probe the deficiencies of existing XC approximations~\cite{Nam2020, Kanungo2021, Kanungo2023}.  

Over the past three decades, several approaches have been developed to numerically solve the inverse DFT problem~\cite{Gorling1992, Wang1993, Zhao1994, Leeuwen1994, Tozer1996, Wu2003, Peirs2003, Jacob2011, Gould2014, Ryabinkin2015, Cuevas2015, Ospadov2017, Jensen2018, Kanungo2019, Shi2021, Stuckrath2021, Shi2022, Gould2023, Aouina2023, Tribedi2023, Kanungo2023}. However, most of the approaches have suffered from numerical artifacts arising from the incompleteness of the underlying basis (e.g., Gaussian) ~\cite{Burgess2007, Bulat2007,Jacob2011} and/or from the incorrect asymptotic behavior of the target densities~\cite{Mura1997, Schipper1997, Gaiduk2013, Kanungo2019}. Two recent approaches have tried to address these challenges using different ideas. The first, hereafter named KZG method, alleviates the numerical challenges by use of a systematically convergent and complete finite-element (FE) basis along with asymptotic corrections to the target density~\cite{Kanungo2019, Kanungo2023}, within a PDE-constrained optimization (PDE-CO) formulation~\cite{Jensen2018, Kanungo2019}. The other, known as the RKS method~\cite{Ryabinkin2015}, makes use of the two-electron reduced density matrix (2-RDM) instead of just the density to find smooth XC potentials, even while working with a finite basis. Conceptually, the KZG method is a \textit{pure density}-to-potential map, as it solely relies on the density, whereas the RKS method is \textit{wavefunction}-to-potential map, as it uses information beyond the density (i.e, the 2-RDM). We remark that although we have characterized the RKS method as an inverse DFT technique in this work, it is often not regarded as one, owing to the fact that it does not guarantee to yield the target density, especially in a finite basis set. However, it is a reliable means to obtain unambiguous and physical XC potentials, and hence, we include it as an inverse DFT method for practical purposes. Although the KZG and RKS methods represent two approaches to inverse DFT, they have their individual strengths and weaknesses. The RKS method, being designed for a finite basis, offers greater efficiency. However, it offers lower accuracy in terms of agreement with the target density, especially when not using a large enough basis. The KZG method offers greater accuracy, but due to the use of a complete FE basis it incurs a higher computational cost. Thus, an accurate and efficient method for inverse DFT is wanting. 

We present a hybrid approach, named KZG-RKS, that combines the RKS and KZG methods to afford both accuracy and computational efficiency. To elaborate, we use the XC potential obtained from the RKS method as initial guess for the KZG method, thereby, significantly accelerating the rate of convergence of its underlying nonlinear optimization. This work also provides, for the first time, a direct comparison of the RKS and KZG, in terms of their XC potentials and agreement with the target densities. We demonstrate the efficacy of the proposed method for several weakly and strongly correlated molecular systems. For all the benchmark systems, we use accurate full configuration interaction (FCI) ground state densities obtained using Slater type orbitals (STOs) instead of the more commonly used Gaussian type orbitals (GTOs). The use of STOs is motivated by their ability to describe the nuclear cusp and the exponential decay in the density. In particular, we employ our recently proposed SlaterRKS~\cite{Tribedi2023}, which modifies the original RKS for STOs so as to enforce the Kato cusp condition~\cite{Kato1957}. Our numerical results demonstrate a substantial $3-11\times$ speedup in the KZG calculations while using the SlaterRKS solution as the initial guess for the XC potential. We observe differences in the XC potentials from SlaterRKS and KZG-RKS near the nuclei and the intershell structure. These differences in the potentials translate to different levels of agreement with the target FCI densities, with the KZG-RKS method providing an order of magnitude better accuracy in the $L_1$ norm of the error in densities.


%% file: methods.tex
\section{Theory and Methods} 
\label{sec:methods}

\subsection{SlaterRKS Method} \label{sec:slaterRKS}

In the formalism proposed by Ryabinkin-Kohut-Staroverov (RKS),\cite{Ryabinkin2015,Cuevas2015} the XC potential is obtained from the 2-RDM of any wavefunction method. Two local energy balance equations (one from wavefunction, the other from Kohn Sham) are compared to evaluate the $\vxc$:
\begin{equation} \label{rks}
    \vxc(\br) = v_{\text{xc},\text{Slater}}^{\text{WF}}(\br) + \frac{\tau^{\text{WF}}(\br)}{\rho^{\text{WF}}(\br)} - \frac{\tau^{\text{KS}}(\br)}{\rho^{\text{KS}}(\br)} + \epsilon^{\text{KS}}(\br) - \epsilon^{\text{WF}}(\br).
\end{equation}
In principle, either GTO or STO basis can be employed to generate the $\vxc$ in the above equation. Ryabinkin et al. demonstrated the stability of the RKS method in obtaining physically accurate XC potentials free from the numerical problems and oscillations which plague other inverse DFT methods performed in finite Gaussian basis sets. However, the densities constructed from GTOs are inaccurate at the nucleus as well as in the long range. Densities and potentials constructed in the STO basis, on the other hand, display correct asymptotic behavior in the long range as well as allows for systematic correction to the nuclear cusp condition through constraints imposed on the occupied MOs.\cite{Schipper1997,handy_molecular_2004} In SlaterRKS,\cite{Tribedi2023} the wavefunction calculation is carried out in a STO basis. In the following, we refer to the reference density obtained from the many-body wavefunction $|\Phi\rangle$ as $\rho^{\text{WF}}(\br)$. The matrix elements of the 2-RDM can thus be calculated as
\begin{equation} \label{2-rdm}
    \Gamma_{pqrs} = \langle \Phi | \hat{a}^\dagger_p \hat{a}^\dagger_q \hat{a}_s \hat{a}_r | \Phi \rangle
\end{equation}
Subsequently, one can define the XC hole density $\rhoWF_{\text{xc}}$ in terms of the 2-RDM and the $\rhoWF$ as
\begin{equation} \label{eq:rhoXC}
    \Gamma(\br,\br_2;\br,\br_2)=\frac{1}{2}\rhoWF(\br)\left[\rhoWF(\br_2)+\rhoWF_{\text{xc}}(\br,\br_2)\right]\,,
\end{equation}
where 
$\Gamma(\br,\br_2;\br',\br'_2) = \sum_{pqrs} \Gamma_{pqrs} \phi^*_p(\br') \phi^*_q(\br_2')\phi_r(\br) \phi_s(\br_2)$
with $\{\phi_p(\br)\}$ being the eigenfunctions of the generalized Fock operator. 
The Slater exchange-correlation charge potential is then evaluated from the pair density:
\begin{equation} \label{vxch}
    v_{\text{xc},\text{Slater}}^{\text{WF}}(\br) = \int \frac{\rhoWF_{\text{xc}}(\br, \br_2)}{|\br - \br_2|} d\br_2.
\end{equation}
The positive-definite kinetic energy density, $\tau^{\text{WF}}(\br)$, and the average local orbital energy, $\epsilon^{\text{WF}}(\br)$, are calculated as,
\begin{equation} \label{wf_KED}
    \tau^{\text{WF}}(\br) = \frac{1}{2}\sum_i n_i |\nabla \phi_i(\br)|^2,
\end{equation}
\begin{equation} \label{wf_ALIE}
    \epsilon^{\text{WF}}(\br) = \frac{1}{\rho^{\text{WF}}(\br)}\sum_i \lambda_i |\phi_i(\br)|^2\,,
\end{equation}
where $\lambda_i$'s are the eigenvalues of the generalized Fock operator and $n_i$'s are their corresponding occupation numbers. For closed-shell systems, as considered in this work, $n_i=2$ for $i \leq N_e/2$ and $0$ otherwise. The corresponding KS terms $\rho^{\text{KS}}(\br)$, $\tau^{\text{KS}}(\br)$ and $\epsilon^{\text{KS}}(\br)$ are evaluated using KS orbitals ($\psi_i$) and corresponding eigenvalues ($\varepsilon_i$), given as 
\begin{equation} \label{ks_rho}
    \rho^{\text{KS}}(\br) = \sum_i n_i |\psi_i(\br)|^2,
\end{equation}
\begin{equation} \label{ks_KED}
    \tau^{\text{KS}}(\br) = \frac{1}{2}\sum_i n_i |\nabla \psi_i(\br)|^2,
\end{equation}
\begin{equation} \label{ks_ALIE}
    \epsilon^{\text{KS}}(\br) = \frac{1}{\rho^{\text{KS}}(\br)}\sum_i n_i\varepsilon_i |\psi_i(\br)|^2.
\end{equation}
Using these $\text{WF}$ and $\text{KS}$ terms, an initial $\vxc$ is generated using Eq. \ref{rks} and the KS eigenvalue problem,
\begin{equation} \label{eq:KSRKS}
    \bigg[ -\frac{1}{2}\nabla^2 + v_{\text{ext}}(\br) + v_{\text{H}}(\br) + \vxc(\br) \bigg] \psi_{i}(\br)=\varepsilon_i\psi_i(\br)\,,
\end{equation}
is solved for a new set of orbitals and eigenvalues. Equations~\ref{ks_rho}, \ref{ks_KED} and \ref{ks_ALIE} are then re-evaluated with the updated $\{\psi_i\}$ and $\{\varepsilon_i\}$ and the process is repeated until the $\vxc$ is self-consistent. Convergence of the SlaterRKS procedure is determined when the $L_1$ norm between the WF density and the KS density remains stable within a tolerance of $10^{-5}$ from one iteration to the next.

The orbitals from both the wavefunction and Kohn Sham formalisms in SlaterRKS are constrained to obey Kato's nuclear cusp condition.\cite{Kato1957} The SCF procedure is therefore modified so the orbitals must satisfy,
\begin{equation} \label{eq:Kato}
    \frac{\partial \psi_i}{\partial \br}\bigg|_{\br = \bR_A} = -Z_A \psi_i(\bR_A),
\end{equation}
where, nucleus $A$ has atomic number $Z_A$ and is at position $\bR_A$.\cite{handy_molecular_2004}

The SlaterRKS method, thus provides XC potentials in finite basis sets from the 2-RDM of a wavefunction without unphysical oscillations. Further, employing Slater basis sets along with the cusp constraint produces more accurate target densities. Overall, this is a practical method to evaluate the XC potential with the promise for incremental advancements towards the exact $\vxc$ through increasing basis set size. However, due to limitations in the availability of larger Slater basis sets, SlaterRKS calculations are presently limited to quadruple zeta basis set quality.

\subsection{KZG Method} \label{sec:KZG}
The KZG method uses the partial differential equation constrained optimization (PDE-CO)~\cite{Jensen2018, Kanungo2019} approach to inverse DFT. For simplicity, we present the PDE-CO formulation assuming non-degenerate KS eigenvalues. However, the numerical implementation of the KZG method used in this work follows the more general formulation that admits degeneracy~\cite{Kanungo2023}. Given a target density $\rhoWF(\br)$ from a many-body wavefunction, the PDE-CO approach to inverse DFT seeks to finds the $\vxc(\br)$ by solving the following optimization problem:
\begin{equation} \label{eq:rhomin}
\text{arg}\min_{\vxc(\br)}\int{w(\br)\left(\rhoWF(\br)-\rhoKS(\br)\right)^2\dr}\,,
\end{equation}
subject to the condition that $\rhoKS(\br)$ is evaluated from the solution of the KS eigenvalue problem and that the KS orbitals are normalized, 
\begin{equation}\label{eq:KS} 
    \left(-\frac{1}{2}\nabla^2+v_{\text{ext}}(\br)+v_{\text{H}}(\br)+\vxc(\br)\right)\psi_{i}(\br)=\epsilon_i\psi_i(\br)\,,
\end{equation}
\begin{equation} \label{eq:Normal}
\int{|\psi_i(\br)|^2\dr} =1\,.
\end{equation}
Note that Eq.~\ref{eq:KS} is same as Eq.~\ref{eq:KSRKS}, albeit the $\vxc$ being from the KZG procedure. In the above equation, $w(\br)$ is an appropriately chosen positive weight to expedite convergence. One can recast the above PDE-CO as an unconstrained minimization of the following Lagrangian:
\begin{equation} \label{eq:L}
\begin{split}
    \mathcal{L} = & \int{w(\br)\left(\rhoWF(\br)-\rhoKS(\br)\right)^2\dr} 
    +\sum_{i=1}^{N_e/2}{\int{p_i(\br)\left(\hat{H}_{\text{KS}}-\epsilon_i\right)\psi_i(\br)\dr}} +\sum_{i=1}^{N_e/2}{\mu_i\left(\int{|\psi_i(\br)|^2\dr}-1\right)}\,,
\end{split}
\end{equation}
where $\hat{H}_{\text{KS}}=-\frac{1}{2}\nabla^2+v_{\text{ext}}(\br)+v_{\text{H}}(\br)+\vxc(\br)$ is the KS Hamiltonian, $p_i$ is the adjoint function that enforces the KS eigenvalue problem for $\psi_i$, and $\mu_i$ is the Lagrange multiplier for the normalization condition on $\psi_i$. Optimizing $\mathcal{L}$ with respect to $p_i$, $\mu_i$ leads to the constraints of Eq.~\ref{eq:KS} and Eq.~\ref{eq:Normal}, respectively. Optimizing $\mathcal{L}$ with respect to $\psi_i$ and $\epsilon_i$ leads to:  
\begin{align} 
(\hat{H}-\epsilon_i)p_i(\br) &= 8w(\br)(\rhoWF(\br)-\rhoKS(\br))\psi_i(\br) - 2\mu_i\psi_i(\br)\,, \label{eq:adjoint} \\
\int{p_i(\br)\psi_i(\br)\dr} &= 0\,. \label{eq:orthoAdjointPsi}
\end{align}
Having solved Eqs.~\ref{eq:KS},~\ref{eq:Normal},~\ref{eq:adjoint},~\ref{eq:orthoAdjointPsi}, the variation of $\mathcal{L}$ with respect to $\vxc$ is given by
\begin{equation} \label{eq:gradVXC}
\frac{\delta\mathcal{L}}{\delta \vxc(\br)} = \sum_{i=1}^{N_e/2}{p_i(\br)\psi_i(\br)}\,.
\end{equation}
The above forms the key equation to update the $\vxc$ via any gradient-based optimization technique. We solve the above set of equations by discretizing the $\psi_i$'s, $p_i$'s, and $\vxc$ using an adaptively refined spectral finite-element (FE) basis~\cite{MOTAMARRI2013308,MOTAMARRI2020106853,das2022dft}. The completeness of the FE basis is crucial to obtaining an accurate solution to the inverse DFT problem. 

We note that although the SlaterRKS procedure enforces the Kato cusp condition on the density at the nuclei, it can contain basis set errors in the density near the nuclei, owing to the incomplete (finite) nature of the Slater basis. As a result, it can create unphysical oscillation in the XC potential near the nuclei. We remedy this by adding a small correction $\Delta \rho(\br)$ to the target density, given as
\begin{equation}\label{eq:deltaRho}
    \Delta \rho(\br) = \rho^{\text{DFA}}_{\text{FE}}(\br) - \rho^{\text{DFA}}_{\text{S}}(\br)\,
\end{equation}
where $\rho^{\text{DFA}}_{\text{FE}}$ denotes the self-consistent groundstate density for a given density functional approximation (e.g., LDA, GGA) that is solved using the FE basis; $\rho^{\text{DFA}}_{\text{S}}$ denotes the same, albeit solved using the Slater basis used in evaluating the target density (i.e., in CI calculation). 
Loosely speaking, $\Delta \rho$, being the difference between two different evaluation of the same physical density (one with a complete FE basis and the other with an incomplete Slater basis), denotes the basis set error in the Slater density, especially near the nuclei. We demonstrate the necessity and the efficacy of the $\Delta\rho(\br)$ correction in Sec.~\ref{sec:deltaRho}.

%% file: results.tex
\section{Results and Discussion} \label{sec:results}
We discuss the various numerical aspects of the proposed approach of combining SlaterRKS and the KZG methods, ranging from the efficacy of the $\Delta \rho(\br)$ correction to the accuracy of the method and the eventual improvement in the computational efficiency.  

\subsection{Computational Details} \label{sec:computeDetails}
We demonstrate the different aspects of our method using various weakly (H$_2$, LiH, H$_2$O, and C$_2$H$_4$) and strongly (stretched and dissociated H$_2$, singlet CH$_2$, and stretched C$_2$H$_4$) correlated molecules. For all the benchmark systems, the target densities and their RDMs (used in SlaterRKS) are obtained using a heat-bath configuration interaction (HBCI) procedure.\cite{Holmes2016,sharma2017,Li2018,Dang2023,Chien2018} Tight thresholds ($10^{-5}$ Ha) were employed to ensure the accuracy of the variational wavefunction for the smaller molecules: H$_2$ and LiH, and slightly relaxed threshold of $10^{-4}$ Ha for CH$_2$, H$_2$O and C$_2$H$_4$. Integrals with Slater basis functions for the HBCI and RKS computations were numerically integrated via the SlaterGPU library,\cite{Dang2022} using an atom-centered grid.\cite{becke_multicenter_1988,mura_improved_1996,murray_quadrature_1993} This three-dimensional grid consists of radial and angular points weighted in accordance with the Becke partitioning scheme. 50 radial and 302 angular points are employed for each atom. The SlaterRKS $\vxc$ is evaluated on the quadrature grid used in the KZG method. Evaluation of terms (Eqs. \ref{vxch}, \ref{wf_KED}, \ref{wf_ALIE}, \ref{ks_KED} and \ref{ks_ALIE}) contributing to the construction of the RKS XC potential on the grid using Eq. \ref{rks} are also accelerated on the GPU using OpenACC.

In SlaterRKS, as $\br\to\infty$, the KS terms in Eq. \ref{rks}, such as $\epsilon^{\text{KS}}(\br)$ and $-\tau^{\text{KS}}(\br)/\rho^{\text{KS}}(\br)$ tend toward the energy of the HOMO of the KS system, $\varepsilon_{\text{HOMO}}$. Similarly, the WF terms, $\epsilon^{\text{WF}}(\br)$ and $-\tau^{\text{WF}}(\br)/\rho^{\text{WF}}(\br)$ approach the first ionization energy derived from the extended Koopman's theorem, $-I_{\text{EKT}}$. It is well-known that $\vxc(\br) \to v_{\text{xc},\text{Slater}}^{\text{WF}}(\br) \approx -1/r$ under this asymptotic limit. To ensure that all terms except $v_{\text{xc},\text{Slater}}^{\text{WF}}(\br)$ cancel in the far field, all KS eigenvalues ${\varepsilon_i}$ are shifted such that $\varepsilon_{\text{HOMO}}=-I_{\text{EKT}}$. Further, to ensure smooth transition of $\vxc(\br)$ to $v_{\text{xc},\text{Slater}}^{\text{WF}}(\br)$ at large distances (i.e. at low density regions where $\rho(\br) < 10^{-5}$), a smoothing function ($F$) is employed such that,
\begin{equation}
    \vxc^{\text{smooth}}(\br) = f(\br)\vxc(\br) + (1-f(\br))v_{\text{xc},\text{Slater}}^{\text{WF}}(\br),
\end{equation}
where,
\begin{equation} \label{eq:smoothf}
    f(\br) = \frac{\rho^{\text{WF}}(\br)}{\rho^{\text{WF}}(\br) + \theta},\quad \text{with~} \theta = 10^{-5}.
\end{equation}

For our HBCI and SlaterRKS calculations, we used the Slater basis developed by Van Lenthe and Baerends~\cite{Van2003}. In particular, we used the QZ4P basis for all the systems, except for C$_2$H$_4$ systems, wherein we used the TZ2P basis. The STO integrals are calculated using the resolution-of-the-identity (RI) approximation. The RI basis is automatically generated by an atom-wise product of basis functions. 

For the KZG calculations, to discretize the KS orbitals ($\psi_i$) and the adjoint functions ($p_i$) we use an adaptively refined fourth-order FE basis. The $\vxc$, being much smoother in comparison, is discretized using linear FE basis. To assess the efficiency gains in KZG-RKS (i.e., KZG method using initial guess from SlaterRKS), we compare it with another KZG calculation, termed KZG-LDA-FA, which employs the initial guess used in earlier work of Kanungo et al.~\cite{Kanungo2019}. In particular, KZG-LDA-FA uses an initial guess for the XC potential that smoothly transitions from the LDA~\cite{Perdew1992} potential in the high density region to the Fermi-Amaldi potential in the low-density region and is given by 
\begin{equation} \label{eq:vxcInitKZGOrig}
    \vxc(\br) = f(\br) \vxc^{\text{LDA}}[\rhoWF](\br) + (1-f(\br))\vxc^{\text{FA}}[\rhoWF](\br)\,, 
\end{equation}
where $\vxc^{\text{LDA}}[\rhoWF]$ and $\vxc^{\text{FA}}[\rhoWF]$ are the LDA and Fermi-Amaldi potentials corresponding to $\rhoWF$, and $f(\br)$ is same as that in Eq.~\ref{eq:smoothf}, albeit with $\theta=10^{-6}$. In order to seamlessly import the XC potential from SlaterRKS into the FE basis, we first evaluate the SlaterRKS potential on a Gauss quadrature grid that is used in the KZG method. Subsequently, we perform an $L_2$ projection to find its representation in the FE basis. In all our KZG calculations, the $L_2$ error in the density---$(\int\left(\rhoWF-\rhoKS\right)^2\dr)^{1/2}$---is driven below $10^{-4}$. We employ the limited-memory BFGS (L-BFGS) algorithm~\cite{Nocedal1980} to solve the nonlinear optimization in KZG. The geometries of all the molecules used in this work are provided in the SI.   

\subsection{Efficacy of $\Delta \rho(\br)$ correction} \label{sec:deltaRho} 
\begin{figure}[htbp]
    \centering
    \includegraphics{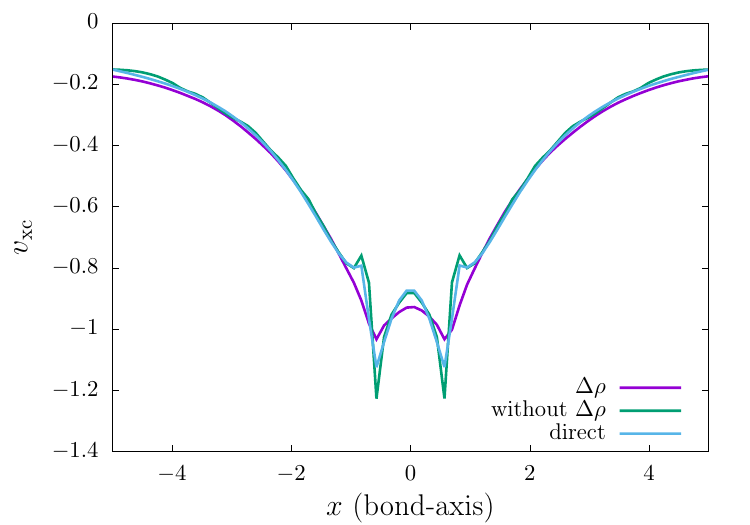}
    \caption{Comparison of the KZG based XC potential for H$_2$(eq) (at equilibrium bond-length) with and without the $\Delta \rho$ correction, along the H-H bond. It also provides a comparison with the direct evaluation of the XC potential, as given in Eq.~\ref{eq:vxcDirect}.}
    \label{fig:H2NoDeltaRho}
\end{figure}

We demonstrate the need for the $\Delta \rho$ correction in alleviating any spurious oscillation in the $\vxc$. To do so, we use the H$_2$ molecule at the equilibrium bond-length $R_{\text{H-H}}= 1.398$ a.u.\---henceforth denoted as H$_2$(eq)---as a benchmark system. The H$_2$(eq) has a single occupied Kohn-Sham orbital, given directly in terms of the target density as $\psi(\br)=(\rhoWF(\br)/2)^{1/2}$. Thus, using Eq.~\ref{eq:KS}, we can directly evaluate the XC potential 
\begin{equation}\label{eq:vxcDirect}
    \vxc^{\text{direct}}(\br)=\epsilon - v_{\text{H}}[\rhoWF](\br) -v_{\text{N}}(\br) + \frac{1}{4}\frac{\nabla^2\rhoWF(\br)}{\rhoWF(\br)} - \frac{1}{8} \frac{|\nabla\rhoWF(\br)|^2}{(\rhoWF(\br))^2}\,,
\end{equation}
where $\epsilon$ is the eigenvalue of the KS HOMO, which for an accurate groundstate $\rhoWF$ is given as $-I_{\text{EKT}}$. In the above expression, given the nuclear cusp in the density, the Laplacian term (fourth term) gradually becomes singular as one approaches the nuclei. For the exact density, this singularity should exactly cancel that of the nuclear potential ($v_{\text{N}}$). While the SlaterRKS procedure ensures the cancellation of the singularities at the nuclei, there still remains basis set errors in the density close to the nuclei due to the incomplete (finite) nature of the Slater basis. As a result, the $\vxc$ exhibits spurious oscillations near the nuclei. We illustrate this in Fig.~\ref{fig:H2NoDeltaRho}, where $\vxc^{\text{direct}}$ exhibits unphysical oscillation near the nuclei, despite the expilcit enforcement of the Kato cusp condition in $\rhoWF$ (see Eq.~\ref{eq:Kato}). Further, the evaluation of $\vxc$ via the KZG method without any $\Delta \rho$ correction also leads to a potential that is very close to $\vxc^{\text{direct}}$. This indicates that the oscillation stems from the basis set errors in the Slater density rather than any numerical aspects of inversion in the KZG approach. However, the use of $\Delta \rho$ correction to $\rhoWF$ alleviates these oscillations, resulting in smooth potentials. 
As an interesting note, if we compare Fig.~\ref{fig:H2NoDeltaRho} to Fig. 4 in Ref.~\cite{Kanungo2019}, wherein the target density was obtained using a large Gaussian basis, it is apparent that oscillations are substantially reduced in the case of Slater density that enforces the Kato cusp condition. This presents an opportunity for a better Slater basis that can dispense the need of $\Delta \rho$ correction.          

\subsection{Comparison of RKS and KZG methods} \label{sec:compare}

In this section, we compare the XC potentials obtained from the RKS and KZG methods, for various benchmark systems. We also measure the speedup afforded by KZG-RKS (i.e., due to the use of the RKS initialization). Additionally, we compare the accuracy attained by KZG-RKS and SlaterRKS, in terms of the error in the density.     

\begin{figure}[htpb] 
    \centering
    \includegraphics{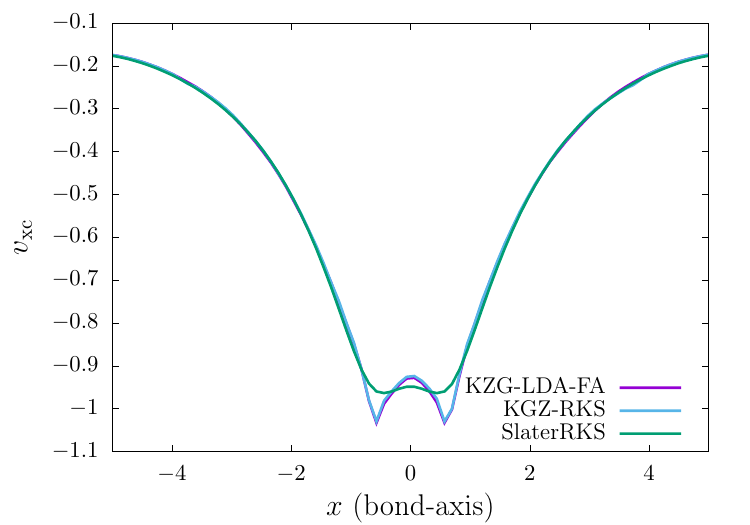}
    \caption{Comparison of the KZG and SlaterRKS based XC potentials for stretched H$_2$(eq), along the H-H bond. KZG-LDA-FA and KZG-RKS refer to the KZG-based solutions obtained using LDA-FA (Eq.~\ref{eq:vxcInitKZGOrig}) and SlaterRKS initialization, respectively.}
    \label{fig:H2_eq}
\end{figure}

Fig.~\ref{fig:H2_eq} compares the KZG-RKS, KZG-LDA-FA, and SlaterRKS based XC potentials for H$_2$(eq). As evident, both the KZG-RKS and KZG-LDA-FA lead to the same XC potential, underlining the robustness of the KZG method with respect to initial guess for the XC potential. However, the KZG and SlaterRKS potentials show differences near the nuclei, where the KZG potential is appreciably deeper than the SlaterRKS potential. 
Notably, as seen from Table~\ref{tab:L1Err}, the KZG approach achieves an order of magnitude better accuracy in the $L_1$ norm error in the density compared to SlaterRKS, owing to the use of a systematically convergent FE basis in the inversion. Further, as noted in Table~\ref{tab:speedup}, we observe a substantial $9\times$ speedup for the KZG-RKS over KZG-LDA-FA, thus highlighting the efficiency and accuracy of the KZG-RKS approach.        

\begin{figure}[htpb] 
    \centering
    \includegraphics{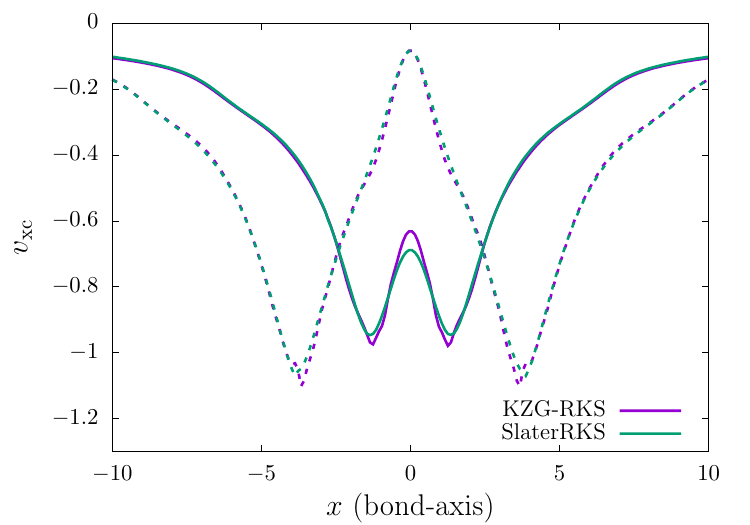}
    \caption{Comparison of the KZG-RKS and SlaterRKS based XC potentials for H$_2$(2eq) and H$_2$(d), along the H-H bond.}
    \label{fig:H2_2eq_H2_d}
\end{figure}

We now present a similar study of other systems. Given that the KZG-LDA-FA and the KZG-RKS lead to the same XC potential, for the subsequent discussion, we only present the results with the KZG-RKS. Fig.~\ref{fig:H2_2eq_H2_d} presents the KZG-RKS and the SlaterRKS XC potentials, respectively, for two stretched H$_2$ molecules: H$_2$(2eq) ($R_{\text{H-H}}=2.83$ a.u., roughly twice the equilibrium bond-length) and H$_2$(d) ($R_{\text{H-H}}=7.56$ a.u., at dissociation). These stretched molecules involve strong electronic correlations, and hence, serve as stringent benchmarks for both KZG and the SlaterRKS. As evident, the XC potentials from both the approaches are qualitatively similar but with notable differences near the nuclei and in the bonding regions. As noted in Table~\ref{tab:L1Err}, for both the systems, the KZG-RKS method attains an order of magnitude better accuracy in the density compared to the SlaterRKS method, attributed to the systematically convergent FE basis in KZG. Once again, the KZG-RKS leads to a significant $3-11\times$ speedup over the KZG-LDA-FA (see Table~\ref{tab:speedup}).       

\begin{figure}[htpb] 
    \centering
    \includegraphics{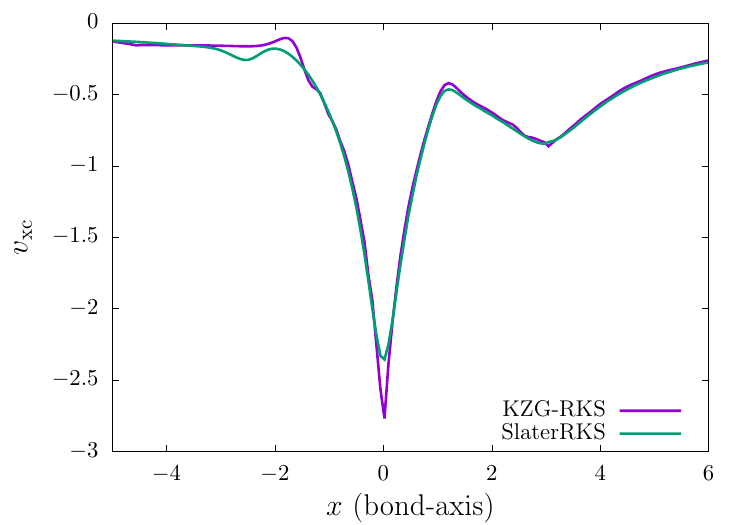}
    \caption{Comparison of the KZG-RKS and SlaterRKS based XC potentials for LiH, along the Li-H bond.}
    \label{fig:LiH}
\end{figure}

\begin{figure}[htpb] 
    \centering
    \includegraphics{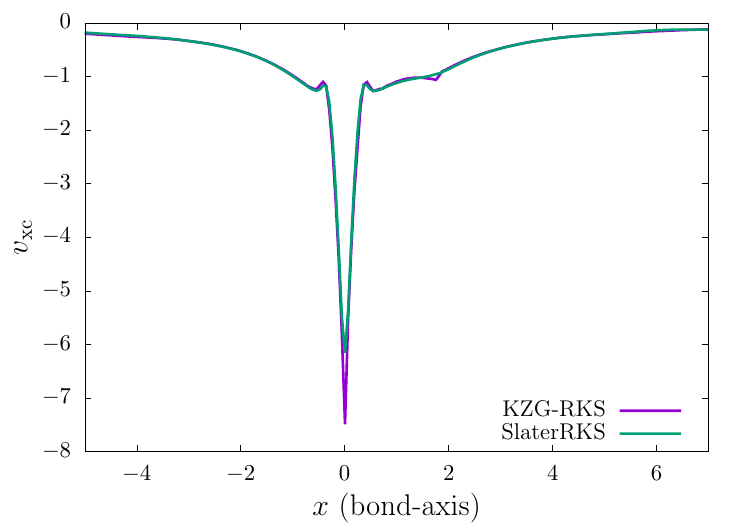}
    \caption{Comparison of the KZG-RKS and SlaterRKS based XC potentials for H$_2$O, along the O-H bond.}
    \label{fig:H2O}
\end{figure}

\begin{figure}[htpb] 
    \centering
    \includegraphics{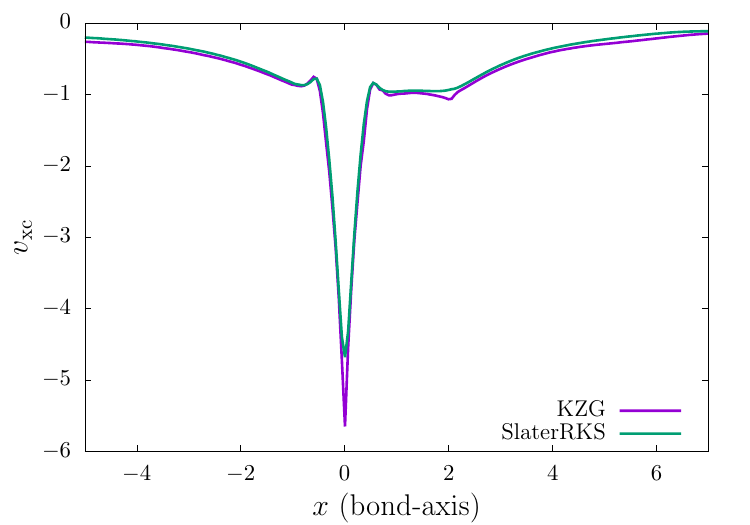}
    \caption{Comparison of the KZG-RKS and SlaterRKS based XC potentials for CH$_2$, along the C-H bond.}
    \label{fig:CH2}
\end{figure}

We now turn to some heteronuclear molecules. Fig.~\ref{fig:LiH}, Fig.~\ref{fig:H2O}, and Fig.~\ref{fig:CH2} show a similar comparison for LiH, H$_2$O, and the singlet state of CH$_2$ radical, respectively. The CH$_2$ radical is a strongly correlated system, offering another challenging benchmark for KZG and SlaterRKS. For both KZG and SlaterRKS potentials, we observe intershell structures near the heavier atom (Li, O, and C atom in LiH, H$_2$O, and CH$_2$, respectively), which is expected in the exact XC potential. Similar to previous examples, while KZG and SlaterRKS potentials have qualitative similarities, they differ appreciably near the nuclei and in the intershell region. Further, as evident from Table~\ref{tab:L1Err}, compared to SlaterRKS, KZG-RKS achieves substantially better accuracy in the density. We attain a $4-7\times$ speedup for KZG-RKS over its KZG-LDA-FA counterpart (see Table~\ref{tab:speedup}).      

\begin{figure}[htpb] 
    \centering
    \includegraphics{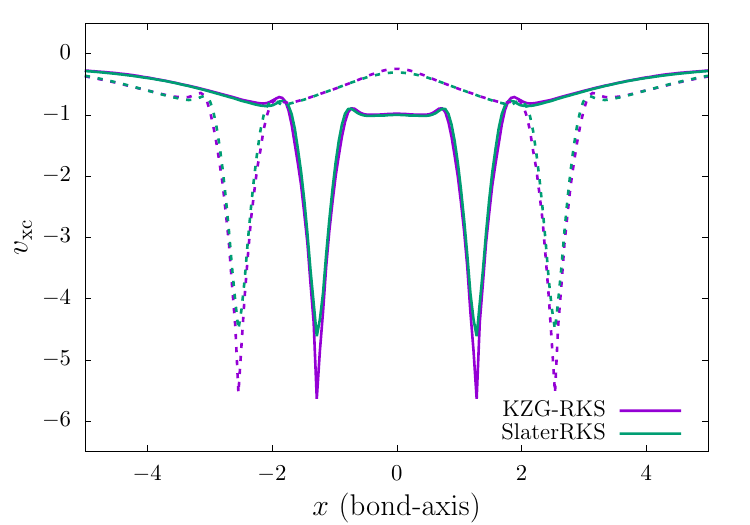}
    \caption{Comparison of the KZG-RKS and SlaterRKS based XC potentials for C$_2$H$_4$(eq) (at equilibrium C-C bond-length), along the C-C bond.}
    \label{fig:C2H4_eq_C2H4_2eq}
\end{figure}

We finally study two polyatomic molcules: C$_2$H$_4$(eq) (at equilibrium C-C bond length, i.e., $R_{\text{C-C}}=2.53$ a.u.) and C$_2$H$_4$(2eq) (at twice the equilibrium C-C bond length, i.e., $R_{\text{C-C}}=5.06$ a.u.). Both the molecules involve double bonds between the C atoms, and hence, are different from the benchmarks considered in our previous SlaterRKS work~\cite{Tribedi2023}. More importantly, the C$_2$H$_4$(2eq) represents a strongly correlated system with a double bond, and these systems have so far received little attention in the context of inverse DFT. As is apparent from Fig.~\ref{fig:C2H4_eq_C2H4_2eq}, both KZG-RKS and SlaterRKS produce qualitatively similar potentials, but with appreciable differences near the nuclei. As shown in Table~\ref{tab:speedup}, the KZG-RKS initialization attains a $3-11\times$ speedup over KZG-LDA-FA. Lastly, as with previous examples, KZG affords significantly better accuracy in the density as compared to SlaterRKS (cf. Table~\ref{tab:L1Err}).  


\begin{table*} [htpb] 
  \caption{Accuracy of KZG-RKS and SlaterRKS in terms of the normalized $L_1$ error in the density, given as $err=\frac{1}{N_e}\int|\rho^{\text{WF}}(\br)-\rho(\br)|dr$.} 
  \label{tab:L1Err}
  \begin{tabular}{| M{3cm} | M{3cm} | M{3cm}|}
    \hline
    \multirow{2}{3cm}{\centering System} & \multicolumn{2}{M{6cm}|}{$err$} \\
    \cline{2-3}
    & KZG-RKS& SlaterRKS \\
    \hline \hline
    H$_2$(eq) & $1.52\times10^{-4}$	& $4.06\times10^{-3}$\\ \hline
    H$_2$(2eq) & $3.36\times10^{-4}$	& $1.37\times10^{-2}$ \\ \hline
    H$_2$(d) & $1.51\times10^{-3}$	& $8.86\times10^{-3}$\\ \hline
    LiH & $5.60\times10^{-4}$	& $4.66\times10^{-3}$\\ \hline
    H$_2$O & $2.94\times10^{-4}$	& $3.28\times10^{-3}$ \\ \hline
    CH$_2$ & $3.16\times10^{-4}$	& $4.91\times10^{-3}$ \\ \hline
    C$_2$H$_4$(eq) & $3.85\times10^{-4}$	& $5.03\times10^{-3}$ \\ \hline
    C$_2$H$_4$(2eq) & $3.49\times10^{-4}$	& $7.62\times10^{-3}$ \\ \hline
    \hline
  \end{tabular}
\end{table*}

\begin{table*} [htpb] 
  \caption{Speedup in KZG method by SlaterRKS initialization. KZG-LDA-FA refers to using initial guess for the XC potential given in Eq.~\ref{eq:vxcInitKZGOrig} and KZG-RKS refers to using the solution of SlaterRKS to initialize the XC potential in the KZG method. $N_{\text{iter}}$ refers to the number of LBFGS iterations taken to reach an accuracy of $10^{-4}$ in $||\rhoWF-\rhoKS||_{L^2} $ in the KZG method.} 
  \label{tab:speedup}
  \begin{tabular}{| M{3cm} | M{3cm} | M{3cm}|}
    \hline
    \multirow{2}{3cm}{\centering System} & \multicolumn{2}{M{6cm}|}{$N_{\text{iter}}$} \\
    \cline{2-3}
    & KZG-LDA-FA & KZG-RKS \\
    \hline \hline
    H$_2$(eq) & 78 & 8\\ \hline
    H$_2$(2eq) & 74 & 22 \\ \hline
    H$_2$(d) & 1210 & 104\\ \hline
    LiH & 183 & 42\\ \hline
    H$_2$O & 902 & 172 \\ \hline
    CH$_2$ & 957 & 136 \\ \hline
    C$_2$H$_4$(eq) & 431 & 149 \\ \hline
    C$_2$H$_4$(2eq) & 866 & 176 \\ \hline
    \hline
  \end{tabular}
\end{table*}

%% file: conclusion.tex
\section{Conclusion} \label{sec:conclusion}
In summary, we have an accurate and computational efficient solution to the inverse DFT problem, named KZG-RKS, by combining the best of the two state-of-the-art inverse Kohn-Sham approaches---Ryabinkin–Kohut–Staroverov (RKS) and the Kanungo-Zimmerman-Gavini (KZG). We used our Slater basis based extension to RKS, named SlaterRKS, to inexpensively evaluate an approximate solution for the XC potential, corresponding to FCI wavefunctions. Subsequently, we used the SlaterRKS solution as an initial guess for the XC potential in the KZG method to accelerate its convergence and obtain a more accurate solution by the use of systematically convergent finite-element (FE) basis. Using several weakly and strongly correlated molecules, we demonstrated a substantial $3-11\times$ speedup for the KZG-RKS method over the original KZG method. Comparing the SlaterRKS and KZG-RKS based XC potentials, while we observed qualitatively similar potentials, there remains notable differences near the nuclei for all systems studied in this work, as well as differences in the intershell and bonding regions for some molecules. Quantitatively, in terms of reproducing the target densities, we observed an order-of-magnitude better accuracy for KZG-RKS over SlaterRKS\cn. We expect the KZG-RKS method to be useful in rapid generation of exact XC potentials from accurate reference FCI wavefunctions, and thereby, aid in understanding and development of better XC functionals.

%% file: bibitem_main.tex
\providecommand{\latin}[1]{#1}
\makeatletter
\providecommand{\doi}
  {\begingroup\let\do\@makeother\dospecials
  \catcode`\{=1 \catcode`\}=2 \doi@aux}
\providecommand{\doi@aux}[1]{\endgroup\texttt{#1}}
\makeatother
\providecommand*\mcitethebibliography{\thebibliography}
\csname @ifundefined\endcsname{endmcitethebibliography}  {\let\endmcitethebibliography\endthebibliography}{}